\documentclass[titlepage,twocolumn,superscriptaddress,prb]{revtex4-1}
\usepackage{amsmath}
\usepackage{amssymb}
\usepackage{graphicx}
\usepackage{color}
\usepackage{floatrow}

\newcommand{\W}{\mathcal{W}}
\newcommand{\I}{\mathrm{I}}
\renewcommand{\c}{\kappa}
\newcommand{\bc}{\bar{\c}}
\newcommand{\SE}{\mathcal{E}}

\begin{document}

\title{Topological quantization of energy transport in micro- and nano-mechanical lattices}

\author{Chih-Chun Chien}

\affiliation{School of Natural Sciences, University of California, Merced, CA
95343, USA}

\author{Kirill A. Velizhanin}

\affiliation{Theoretical Division, Los Alamos National Laboratory, Los Alamos,
NM 87545, USA}

\author{Yonatan Dubi}

\affiliation{Department of Chemistry and the Ilse Katz Institute for Nanoscale
Science and Technology, Ben-Gurion University of the Negev, Beer-Sheva
84105, Israel}

\author{B. Robert Ilic}

\affiliation{Center for Nanoscale Science and Technology, National Institute of
Standards and Technology, Gaithersburg, MD 20899, USA}

\author{Michael Zwolak}

\email{mpz@nist.gov}

\affiliation{Center for Nanoscale Science and Technology, National Institute of
Standards and Technology, Gaithersburg, MD 20899, USA}

\begin{abstract}
	Topological effects typically discussed in the context of quantum physics are emerging as one of the central paradigms of physics. Here, we demonstrate the role of topology in energy transport through dimerized micro- and nano-mechanical lattices in the classical regime, i.e., essentially ``masses and springs''. We show that the thermal conductance factorizes into topological and non-topological components. The former takes on three discrete values and arises due to the appearance of edge modes that prevent good contact between the heat reservoirs and the bulk, giving a length-independent reduction of the conductance. In essence, energy input at the boundary mostly stays there, an effect robust against disorder and nonlinearity. These results bridge two seemingly disconnected disciplines of physics, namely topology and thermal transport, and suggest ways to engineer thermal contacts, opening a direction to explore the ramifications of topological properties on nanoscale technology.	
\end{abstract}

\maketitle

\section{Introduction}

Topology gives rise to fascinating phenomena and can lead to the emergence of many exotic states of matter~\cite{Hasan10,Qi11,Bernevig_book,ShenBook}, from condensed matter~\cite{Chiu2016,Bansil2016} to cold atoms~\cite{Goldman2016} to quantum computation~\cite{Nayak08,Stern2013}. 
An example of a lattice with a non-trivial topology is the Su-Schrieffer-Heeger (SSH) model of electrons hopping in polyacetylene~\cite{SSH79}, which is the focus of many cold-atom studies, e.g., for measuring the Zak phase~\cite{atala_direct_2013} and demonstrating topological Thouless pumping~\cite{nakajima_topological_2016,lohse_thouless_2016}. 
While there are works focusing on topological effects in the classical regime~\cite{Kane14,Chen14,Susstrunk15,Paulose15,Huber16,Lu14,Berg11,Yang15,Kim17,Wang15,Khanikaev15,Ong16}, few connect topology and energy transport~\cite{Bid10}. 

\begin{figure*}[t]
\begin{centering}
\includegraphics[width=0.85\textwidth]{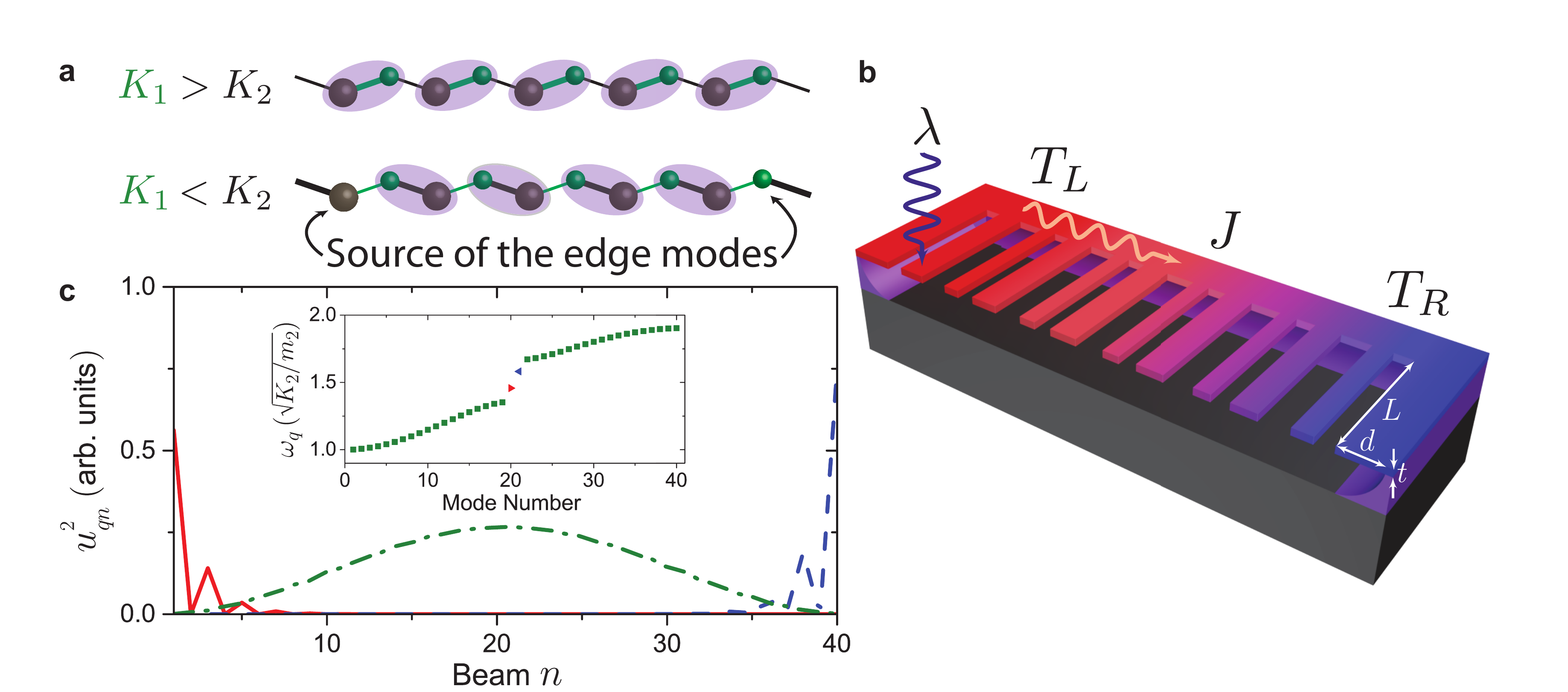}
\par\end{centering}
\caption{
(a) Illustration of dimerization in a mechanically alternating lattice.
Intracell ($K_1$) and intercell ($K_2$) nearest-neighbor couplings are shown by green and black lines, respectively. Depending on the value of the winding number, Eq.~(\ref{eq:wn}), and the parity of the lattice length, $N$, there can be zero, one, or two edge modes, Eq.~(\ref{eq:Nes}).
For example, when $N$ is even and $K_{1}>K_{2}$ then $\W=0$ ($K_{1}<K_{2}$ with $\W=1$) and no edge mode (two edge modes) are present as the strong
couplings pair all the sites (all but the two sites at the boundaries) as shown in the left (right) lattice.
(b) Illustration of energy transport in an envisioned micro- or nano-mechanical topological lattice. The large and small spacings between beams of length $L$ alternate the nearest neighbor couplings (varying widths can alternate other parameters). Other device characteristics (thickness $t$, undercut $d$, device materials) can be used to tune the parameters in Eq.~\eqref{eq:Hq}. A modulated laser of wavelength $\lambda$ can equilibrate the end beam of the lattice at some elevated temperature $T_{L}$, while the other end is either damped into equilibrium with its surroundings at $T_{R}$ or its local temperature is measured optically or electronically via its oscillations. A difference in these two temperatures will drive an energy current $J$. The presence of the edge modes will create an interfacial resistance, as the localized modes tend to decouple the bulk lattice from the boundaries, thereby reducing the ability of energy to flow away from the edge.
(c) Three representative normal modes, plotted as the polarization vector squared, $u_{qn}^2$, versus beam position, $n$, along the lattice. The two edge modes (solid red and dashed blue lines) are localized around the left and right edges, respectively, while delocalized modes are spread across the entire lattice (dot-dashed green line shows $\sqrt{N} u^2_{qn}$ for one delocalized mode). The parameters are $K_2=2 K_1$, $m_2=3 m_1/4$, and $\omega_2=\omega_1$. The inset shows the frequencies of all modes enumerated in ascending order of frequency. The edge modes reside in the gap between the two bands of delocalized modes.}
\label{fig:Schematic}
\end{figure*}

We present a mechanical system that manifests topological effects in energy transport and has relevance to many nanoscale scenarios~\cite{dhar2008,Xu2016}. This system is the mechanical counterpart to the SSH model in Fig.~\ref{fig:Schematic}(a) where alternating nearest neighbor coupling strengths ``dimerize'' the lattice. When both ends terminate on weak bonds, the whole lattice pairs into dimers. Terminating on a strong bond, though, leaves the end sites unpaired, resulting in the formation of an edge mode. Hence, depending on the topology -- e.g., swapping the nearest neighbor couplings constants, which does not change the bulk -- there will be zero, one (on either the left or right), or two edge modes.

%
%

\section{Results}

The mechanical lattice we examine has vibrational spectrum equivalent to the energy spectrum of the SSH model when all parameters (except the alternating nearest-neighbor coupling) are uniform. Otherwise it is identical to the spectrum of a slice of the time-dependent Rice-Mele model, which has a quantized Chern number on the extended 2D plane~\cite{ShenBook}. The lattice has the Hamiltonian
$H=\sum_{n}\frac{m_{n}}{2}\left(\dot{x}_{n}^{2}+\omega_{n}^{2}x_{n}^{2}\right)+\sum_{n}\frac{K_{n}}{2}(x_{n}-x_{n+1})^{2}$
with masses $m_{n}$, onsite frequencies $\omega_{n}$, and nearest-neighbor couplings $K_n$ for site $n$ with coordinate $x_n$. These parameters, $[m_n,\omega_n,K_n]$, are $[m_1,\omega_1,K_1]$ or $[m_2,\omega_2,K_2]$ for odd or even $n$, respectively. After a lattice Fourier transform, we get the Bloch Hamiltonian
\begin{equation}
{\bf H}_{q}=h_{0}\I+h_{z}\sigma_{z}+{\bf \bar{H}}_{q},\label{eq:Hq}
\end{equation}
where $\sigma_z$ is the $z$ Pauli matrix, $\I$ is the $2\times 2$ identity matrix, and $h_z$ and $h_0$ are given in the Supplemental Material (SM)~\cite{SI}, and
\begin{equation}
{\bf \bar{H}}_{q}=-\frac{1}{m}\begin{bmatrix}0 & f^*(q)\\
f(q) & 0
\end{bmatrix} , \label{eq:Hqbar}
\end{equation}
with $m=\sqrt{m_1 m_2}$ and $f(q)=K_1+K_2 e^{iq}$. A natural realization of this model is in one-dimensional micro- and nano-electromechanical systems (MEMS-NEMS), which provide a versatile platform for dynamical phenomena and devices~\cite{Sato2006}. As we will discuss, a combination of laser-induced heating and optical/electronic readout can topologically characterize energy transport in MEMS-NEMS, as shown in Fig.~\ref{fig:Schematic}(b). However, since this model is one of the most elementary examples of a physical system -- classically coupled ``masses and springs'' -- there exists many alternative realizations.

\begin{figure}
\begin{centering}
\includegraphics[width=1\columnwidth]{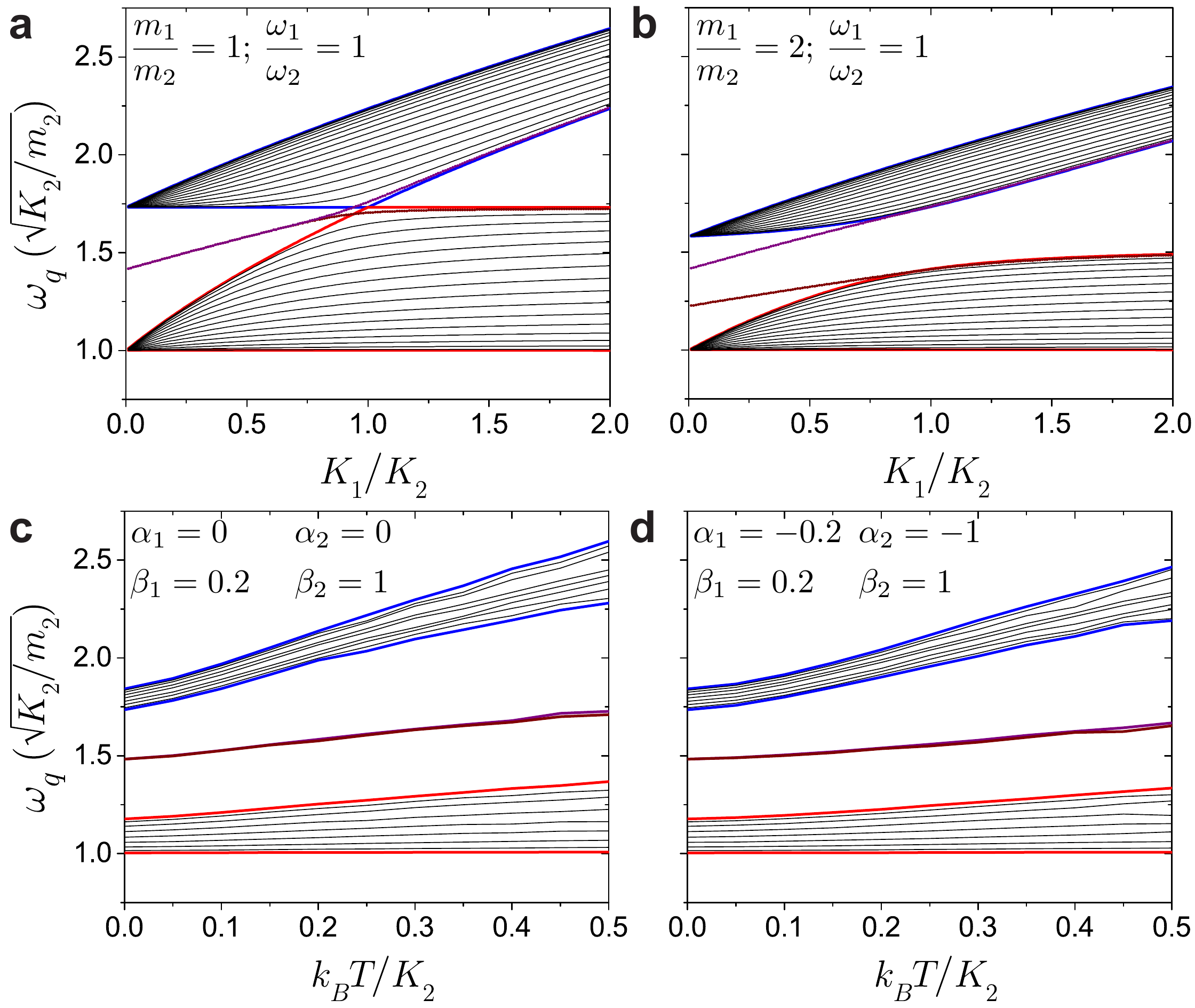}
\par\end{centering}
\caption{Band structure and edge modes for (a,b) the harmonic lattice (that leads to Eq.~\eqref{eq:Hq}) versus the ratio of couplings and (c,d) a nonlinear, Fermi-Pasta-Ulam (FPU) like~\cite{Burkalov1990,James2007,Livi1997,Aoki1992} generalization versus temperature. (a,b) When $K_{1}<K_{2}$ in an even site lattice,
there are two edge modes (purple and wine lines) that reside in the gap between the two phonon
bands (outlined with blue and red for the upper and lower bands, respectively). As $K_{1}$ increases the spatial extent of the edge
states grows until they merge with the bulk
states at $K_{1}=K_{2}$. This process is shown for 
alternating couplings only (i.e., $m_{1}/m_{2}=1$; $\omega_{1}/\omega_{2}=1$)
and for the masses alternating (i.e., $m_{1}/m_{2}=2$; $\omega_{1}/\omega_{2}=1$). (c,d) The edge modes also exist in nonlinear lattices and  persist even as the nonlinearity increases with temperature $T$ (at high enough temperatures, the nonlinearity merges the edge modes with the bulk).
The persistence of the edge modes is relevant to MEMS/NEMS, which are often operated in nonlinear regimes~\cite{Sato2003,Sato2006}. Prior studies also show that classical nonlinear systems can exhibit topological excitations~\cite{Chen14}.
\label{fig:States}}
\end{figure}

The topological nature of the lattice can be seen by considering ${\bf \bar{H}}_{q}=R_x\sigma_x+R_y\sigma_y$, where $\sigma_{x,y}$ are the Pauli matrices. The curve $(R_x=K_1+K_2\cos(q),R_y=-K_2\sin(q))$ may or may not wrap around the origin in the complex plane as $q$ goes from 0 to $2\pi$. Counting how many times the curve encircles the origin gives the {\it winding number},
\begin{equation}
\W=\begin{cases}
1, & K_{1}<K_{2}\\
0, & K_{1}>K_{2}
\end{cases} . \label{eq:wn}
\end{equation}
This number is an important topological property of a 1D system's band structure~\cite{Hasan10,Chiu2016}. The Zak phase~\cite{Zak89} is the 1D Berry phase and is $2\pi$ times the winding number. When the winding number is nonzero, the lattice is topologically non-trivial and edge modes appear, decaying exponentially from the edges with a decay length $\xi=-\ln (K_1/K_2)$. Without loss of generality, we use the convention that if only one edge mode is present, it is on the left. The number of left ($N_L$) and right ($N_R$) edge modes is thus
\begin{align}
N_L&=\W,~ N_R=\frac{1-e^{i\pi(N+\W)}}{2}.\label{eq:Nes}
\end{align}
Figure~\ref{fig:Schematic}(c) shows the decaying amplitude squared of two edge modes.  The $\sigma_z$ term in Eq.~\eqref{eq:Hq}, while not present in the SSH model, does not destroy the edge modes as one can verify explicitly (see the SM~\cite{SI}). Moreover, the edge modes persist in the presence of nonlinearity (see Fig.~\ref{fig:States}).

These modes are more than just a physical curiosity, however. Figure~\ref{fig:Schematic}(b) illustrates a lattice of interacting cantilevers with one end at a temperature $T_L$. When the lattice is locally excited at the boundary, e.g., via a laser as shown in Fig.~\ref{fig:Schematic}(b), or via an electromagnetic coupling, the resulting energy current will depend on the presence of the edge modes, whether this energy flow is due to a single transient excitation or in a steady-state. The conductance, $\c \equiv J/\Delta T$, where $\Delta T$ is the temperature difference between two reservoirs, captures this effect. For the bulk, the intrinsic conductance, $\c_{0}$, is given by its average phonon group velocity \cite{velizhanin15}
\begin{equation}
\c_{0}=\frac{k_{B}}{2\pi}\int_{\Omega}dq\,v_{q}=\frac{k_{B}\Omega}{2\pi},\label{eq:Intrinsic}
\end{equation}
where we use $\Omega$ to indicate both the bands and the total bandwidth.
This is the maximum rate at which a harmonic lattice can transport heat
between two equilibrium reservoirs at different temperatures. Since it depends only on the bulk band structure, $\Omega$, it is independent of winding number, i.e., swapping the order of $K_1$ and $K_2$ -- or changing the parity of the lattice -- will not affect it. Reaching
this conductance in practice, however, requires that all phonon modes are sufficiently
in contact with the reservoirs so that they are supplied ample thermal
energy \cite{velizhanin15}. In the presence of topological edge modes this limit is never reached, and the thermal conductance is always lower than $\kappa_0$, regardless of the system length. This is rather surprising, considering the fact that there are at most two edge modes, whereas the number of modes grows linearly with the system size.

We note that the ability of a specific mode $q$ to conduct heat will
depend on its contact with the external reservoirs and its intrinsic
conductance (determined by its group velocity). In the setup of Fig.
\ref{fig:Schematic}(b), the strength of the contact of a specific mode $q$ with the reservoirs
is given by $\gamma u_{q1}^{2}$ and $\gamma u_{qN}^{2}$ for the
left and right, respectively. The coupling (i.e., damping
rate) $\gamma$ is the strength of contact of the reservoirs to the cantilever beam at the
lattice boundary. The polarization vector of the mode on the boundaries,
$u_{qn}^{2}$ with $n=1$ or $N$, attenuates the coupling of the mode $q$ to the reservoirs. When the lattice weakly contacts the reservoirs -- in order to minimally perturb the boundaries -- the conductance for mode $q$,
$\c_{q}$, is due to two contributions in series (see the SM~\cite{SI})
\begin{equation}
\frac{k_{B}}{\c_{q}}=\frac{1}{\gamma u_{q1}^{2}}+\frac{1}{\gamma u_{qN}^{2}},\label{eq:SeriesCond}
\end{equation}
where the first term is from the left interface and the second from the right interface.
To describe the behavior for arbitrary $\gamma$, the bulk contribution -- $N/v_{q}$, the intrinsic ability of the mode to transfer
heat -- and an overdamping contribution proportional to $\gamma$ would need to be included in Eq.~\eqref{eq:SeriesCond}.
Many of the results below hold up to moderate values of $\gamma$, as explained in the SM~\cite{SI}.

The edge modes have an exponentially vanishing amplitude,
$u_{qn}^{2}\approx0$, for either $n=1$ or $N$, which yields $\c_{q}\approx0$
for $q\in\SE$, where $\SE$ is the set of edge modes. The total conductance will then be
\begin{equation}
\c=\frac{1}{2\pi}\int_{\Omega}dq\,\lim_{N\to\infty}N\c_{q},\label{eq:Cond}
\end{equation}
where the integral is over only the phonon bands and thus the edge
state contribution -- which would be a separate sum -- is absent.
This equation has a similar form to Eq. \eqref{eq:Intrinsic} but
$\c_{q}$ contains the non-ideal contact to the
external heat source and sink.

We proceed by giving a heuristic derivation of the effect of topology, and a rigorous derivation is in the SM~\cite{SI}.
Considering all normal modes of a lattice, one has simple ``sum rules''
for the boundary amplitudes, $\sum_{q}u_{q1}^{2}=1/m$ and $\sum_{q}u_{qN}^{2}=1/m$
for the case when $m_{1}=m_{N}=m$, that reflect the (mass) scaling
and orthogonal transformations that yield the normal modes. In the absence
of edge modes, the bulk modes have a contact strength $u_{qn}^{2} \propto \gamma/(mN)$ for
$n=1$ and $N$.
Using this value for $u_{qn}^{2}$, the non-topological interfacial conductance for an even length lattice is
\begin{equation}
\bc = \frac{k_{B}\gamma}{2m},\label{eq:CondBand}
\end{equation}
which is limited by the coupling of the external reservoirs to the lattice, i.e., the
heat injected is the bottleneck to current flow \cite{velizhanin15} (a similar situation occurs in electronic transport~\cite{Gruss2016,elenewski2017master,gruss2017limits}).

In the presence of edge modes -- states localized at the boundaries
-- the total coupling of the bulk to the reservoirs is reduced: $\sum_{q\in\Omega}u_{qn}^{2}=1/m-\sum_{q\in\SE}u_{qn}^{2}$.
The bulk modes therefore have a contact strength $\propto \gamma\left(1-m\sum_{q\in\SE}u_{qn}^{2}\right)/(mN)$.
The amplitude squared of an edge mode on a boundary of its origin
is $(1-e^{-2\xi})/m$, which follows from the normalization of
an exponentially decaying state (see the SM~\cite{SI}). The bulk modes therefore have contact $\gamma u_{qn}^{2} \propto \gamma\exp\left(-2\xi\right)/(mN)$ for
$n=1$ and $N$. The conductance in the presence of  edge modes is then
\begin{equation}
\c = \frac{k_{B}\gamma e^{-2\xi}}{2m} = e^{-2\xi}\bc.\label{eq:CondWEdge}
\end{equation}
Thus, there is a topologically induced component of the conductance that manifests itself as a prefactor $e^{-2\xi}$.

The case of odd or even $N$ can support $N_L=0,1$ states on the left and and $N_R=0,1$ states on the right, according to the winding number, Eq. \eqref{eq:Nes}. Generalizing Eq.~\eqref{eq:CondWEdge} to arbitrary length lattices, and also inhomogeneous mass and on-site frequency cases, the conductance of the lattice is
\[
\c = \varXi \bc,\label{eq:TopoCond}
\]
where $\bc$ is the nontopological component of the conductance (for $N$ odd, $\bc = k_B\gamma/2 m_1$; for $N$ even, $\bc$ is the conductance in the absence of edge modes, i.e., $K_1$ and $K_2$ swapped, see the SM~\cite{SI}) and $\varXi$ gives the three discrete topological levels
\[
\varXi = \frac{2}{e^{2 N_L \xi}+e^{2 N_R \xi}}. \label{eq:TopoLevels}
\]
The quantity $\varXi$ is thus a function of winding number, as $N_{L(R)}$ depend on it through Eq. \eqref{eq:Nes}.
Out of the different configurations (using the parameters that give the same bulk properties but ordering them in different ways), Eq. \eqref{eq:TopoLevels} will give only three possible values, corresponding to the presence of 0,1 or 2 edge modes. All trivial mass effects (at the boundaries) are in $\bc$.

\begin{figure*}[th]
	\begin{centering}
		\includegraphics[width=1\textwidth]{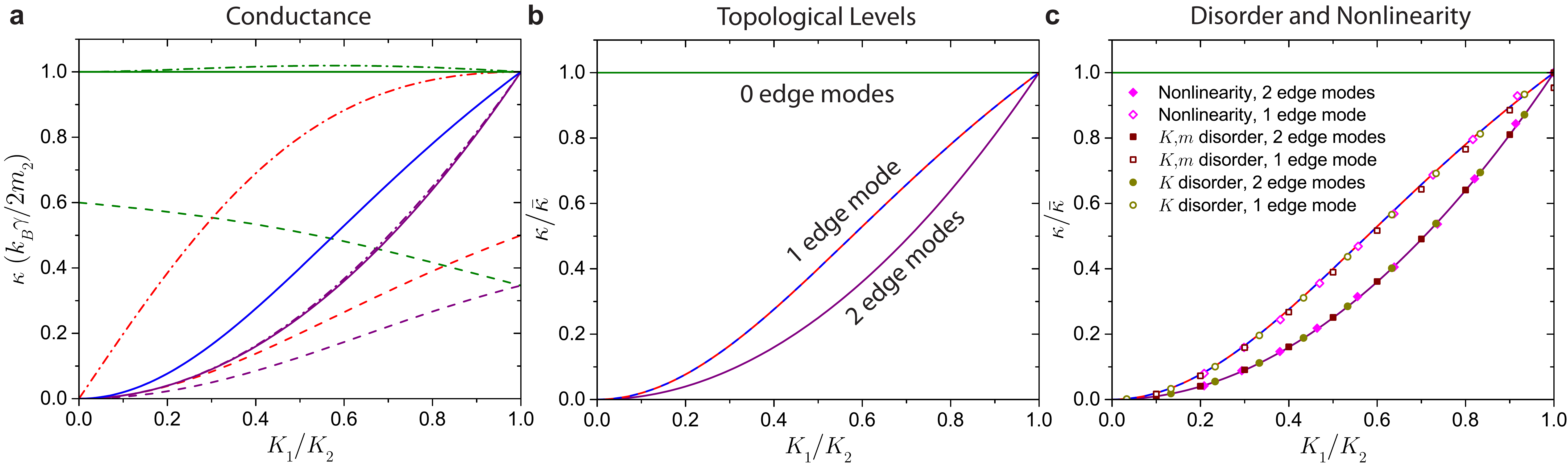}
		\par\end{centering}
	\caption{Topological quantization of the conductance.
		(a) The conductance versus $K_{1}/K_{2}$ of infinite length dimerized
		lattices (in the setup shown in Fig.~\ref{fig:Schematic}(b)). We show only a subset of the curves for clarity, see the SM~\cite{SI} for details. (b) When taking the ratio of the conductance to the non-topological component, $\bc$, all the curves in (a) collapse onto three distinct levels. Green, purple, and red-blue lines show the cases with zero, two, and one edge mode(s) (left or right), respectively, generically suppressing the conductance. (c) The topological conductance is robust to nonlinearity and disorder. The yellow circles and wine squares show the normalized conductance of two disordered lattices (with both an odd and even number of sites), and the magenta diamonds show the normalized conductances of the FPU lattice. Empty (full) symbols are for one (two) edge modes. Since the nonlinearity also introduces an overall shift in the parameters, the data points are plotted versus an effective $K_{1}/K_{2}$, as described in the SM~\cite{SI}. In all cases, the results agree with the theoretical result, Eqs.~\eqref{eq:TopoCond} and \eqref{eq:TopoLevels}.
		\label{fig:ThermalCurrent}}
\end{figure*}

\section{Discussion}

The effect of topological edge modes on thermal conductance is demonstrated in Figure~\ref{fig:ThermalCurrent}, which discusses a uniform lattice with alternating $K_1$ and $K_2$ only (solid lines), a lattice with $m_1=2m_2$ and $\omega_1=\omega_2$ (dashed lines) and a lattice with $m_1/m_2=K_1/K_2,~\omega_1=\omega_2$ (dash-dotted lines). For each bulk lattice, we show the case with zero edge modes (green), two edge modes (purple), and one edge mode (red and blue for a left and right edge mode, respectively).

Figure \ref{fig:ThermalCurrent}(a) plots the thermal conductance $\c$ versus the ratio $K_{1}/K_{2}$. The conductance can take on essentially any value (by changing masses and on-site frequencies, one can fill in the whole plot). However, when taking simple ratios, $\c / \bc$, of the thermal conductance, a simple quantization emerges, as shown in Fig. \ref{fig:ThermalCurrent}(b). These ratios take on just three values given by $\varXi$. Generically, the introduction of edge modes suppresses the conductance, as it reduces the contact between the energy sources/sinks and the bulk states. Non-topological effects (i.e., the changing bulk state structure as $K_1/K_2$ increases) can also significantly influence the conductance for certain sets of parameters~\cite{ChienExactDiag16}.

Since the suppression of the thermal conductance is a topological effect, it will not depend on the specific details of the system and reservoirs and is anticipated to be robust against various modifications to the lattice (so long as the topology is maintained). We demonstrate this robustness in Fig.~\ref{fig:ThermalCurrent}(c) where we plot the normalized conductance versus $K_1/K_2$ for three additional lattices; the FPU-$\beta$ lattice of Fig.~\ref{fig:States}(c) and two disordered, dimerized lattices. For all these cases, the conductance follows Eq.~\eqref{eq:TopoCond}, showing the universality of the topology-induced reduction of thermal conductance. If the Langevin reservoirs are replaced by uniform harmonic lattices with constant coupling $0 < K < \min(K_1,K_2)$ representing trivial topology, the edge modes and their influence on thermal conductance should still survive at the boundary due to a change of topology.

We further note that Eq.~\eqref{eq:SeriesCond} is a general result. It entails, therefore, that even non-topological localized modes (e.g., due to a light mass at the boundary) can suppress the thermal conductance. However, non-topological modes will not display the quantized conductance of Eq.~\eqref{eq:TopoCond} and shown in Fig.~\ref{fig:ThermalCurrent}b,c. As well, there are many channels for heat/energy transport. In the setup envisioned in Fig.~\ref{fig:Schematic}, heat will also be carried by vibrations of the underlying crystal lattice. Therefore, it is necessary to use a low thermal conductivity material so that vibrations of the cantilevers are the dominant channel for energy transport.

\section{Conclusion}

Just as thermal transport can serve
as a probe of nonlinear structural transitions \cite{VelizhaninPRE11,ChienNanotech13}, these results show that signatures
of nontrivial topologies appear in classical (or quantum) energy transport in conventional
physical systems, such as MEMS/NEMS or at crystal-polymer interfaces.  
In particular,  a combination of laser-induced heating and optical/electronic readout will allow for the topological characterization of energy transport in micro- or nano-mechanical lattices and control of heat flow \cite{Li2012}.
The emergence of edge states may help design,
e.g., thermoelectric devices, where the lattice thermal conductance
needs to be suppressed independently of the electronic conductance.  
Moreover, energy flow and thermal properties are critical to the operation of nanotechnologies, where they can limit and even define the functionality of devices~\cite{Dubi11-1,Li2012,Yang2012}.
The results presented here thus generate exciting prospects for observing topological properties in conventional physical systems and utilizing them to design micro- and nano-scale devices. \\


\section*{ACKNOWLEDGMENT}

K.A.V. was supported by the U.S. Department of Energy through the LANL/LDRD Program. Y.D. acknowledges support from the Israel Science Fund (Grant No. 1256/14).

\bibliography{./reference1} 

\end{document}


\title{Topological quantization of energy transport in micro- and nano-mechanical lattices -- Supplemental Information}

\author{Chih-Chun Chien}

\affiliation{School of Natural Sciences, University of California, Merced, CA
95343, USA}

\author{Kirill A. Velizhanin}

\affiliation{Theoretical Division, Los Alamos National Laboratory, Los Alamos,
NM 87545, USA}

\author{Yonatan Dubi}

\affiliation{Department of Chemistry and the Ilse Katz Institute for Nanoscale
Science and Technology, Ben-Gurion University of the Negev, Beer-Sheva
84105, Israel}

\author{B. Robert Ilic}

\affiliation{Center for Nanoscale Science and Technology, National Institute of
Standards and Technology, Gaithersburg, MD 20899, USA}

\author{Michael Zwolak}

\email{mpz@nist.gov}

\affiliation{Center for Nanoscale Science and Technology, National Institute of
Standards and Technology, Gaithersburg, MD 20899, USA}

\maketitle
\tableofcontents
\clearpage

\section{Method} \label{A7}
Thermal conductance calculations were performed by attaching the real-space Hamiltonian (defined above Eq. (1) in the main text) to Langevin reservoirs at the terminating lattice sites. Analytic and numerical calculations follow from deriving the Newton-Langevin equations. For the FPU chain, self-consistent phonon theory is used to address non-linearity. All the details of our numerical and analytical calculations, including rigorous derivations of the equations in the main text, are summarized in the following sections.

\section{Band Structure} \label{A1}

The Hamiltonian of the infinite dimerized lattice is
\begin{equation}
H=\sum_{n=1}^{N}\frac{m_{n}}{2}\left(\dot{x}_{n}^{2}+\omega_{n}^{2}x_{n}^{2}\right)+\sum_{n=0}^{N}\frac{K_{n}}{2}(x_{n}-x_{n+1})^{2},\label{eq:H}
\end{equation}
with $n=0,\pm1,\pm2,...$. Here, $m_{n}$ is the mass at site $n$, $\omega_{n}$ is the frequency of the on-site binding harmonic potential, and $K_n$ is the nearest-neighbor coupling constant. The Hamiltonian parameters $[m_n,\omega_n,K_n]$ are set to either $[m_1,\omega_1,K_1]$ or $[m_2,\omega_2,K_2]$ depending on whether $n$ is odd or even, respectively. For definiteness, the unit cell is taken to be sites 1 and 2, so that  $K_{1(2)}$ is the intracell (intercell)  coupling constant. This gives the equations of motions 
\begin{eqnarray}\label{eq:EOM1}
m_n\ddot{x}_n&=&-(\omega_n^2+K_{n-1}+K_{n})x_n+ K_{n-1}x_{n-1}+K_{n}x_{n+1}.
\end{eqnarray}
The 0$^\mathrm{th}$ and $(N+1)^\mathrm{th}$ oscillators are fixed.

The full Hamiltonian with alternating coupling constants, masses, and on-site
frequencies is invariant with respect to the two-site translation,
$n\rightarrow n+2$. Under this condition, its eigenfunctions have
a Bloch wave form.
Specifically, we perform the standard substitution: (\emph{i}) $\ddot{x}_n\rightarrow -\omega^2_q x_n$, (\emph{ii}) $x_{2l+1}=y_{1}e^{iql}$ and $x_{2l+2}=y_{2}e^{iql}$, where $l=0,\pm1,\pm2,...$ enumerates 2-site unit cells and $q$ is the dimensionless quasimomentum. The lattice Fourier transform reduces the
full problem, Eq.~(\ref{eq:EOM1}), to the $2\times2$ matrix eigenvalue problem 
\begin{widetext}

\begin{equation}
\begin{bmatrix}\omega_{1}^{2}+(K_{1}+K_{2})/m_{1} & -(K_{1}+K_{2}e^{-iq})/m_{1}\\
-(K_{1}+K_{2}e^{iq})/m_{2} & \omega_{2}^{2}+(K_{1}+K_{2})/m_{2}
\end{bmatrix}\begin{pmatrix}y_{1}\\
y_{2}
\end{pmatrix}=\omega_{q}^{2}\begin{pmatrix}y_{1}\\
y_{2}
\end{pmatrix}.
\end{equation}
Using mass-weighted coordinates, $u_{i}=\sqrt{m_{i}}y_{i}$, and rearranging gives
\begin{equation}
\begin{bmatrix}\omega_{1}^{2}+(K_{1}+K_{2})/m_{1}-\omega_{q}^{2} & -(K_{1}+K_{2}e^{-iq})/\sqrt{m_{1}m_{2}}\\
-(K_{1}+K_{2}e^{iq})/\sqrt{m_{1}m_{2}} & \omega_{2}^{2}+(K_{1}+K_{2})/m_{2}-\omega_{q}^{2}
\end{bmatrix}\begin{pmatrix}u_{1}\\
u_{2}
\end{pmatrix}=0,
\label{Eq:DynMat}
\end{equation}
where the matrix is Hermitian. Straightforward diagonalization of this matrix results in two dispersive bands, \begin{equation}
\omega_{\pm}=\sqrt{\frac{1}{2}\left(b_{1}+b_{2}\pm\sqrt{(b_{1}-b_{2})^{2}+4(d_{1}^{2}+d_{2}^{2}+2d_{1}d_{2}\cos(q))}\right)}.
\end{equation}
Here, $b_{1}=(\omega_{1}+K_{1}+K_{2})/m_{1}$, $b_{2}=(\omega_{2}+K_{1}+K_{2})/m_{2}$,
$d_{1}=K_{1}/\sqrt{m_{1}m_{2}}$, and $d_{2}=K_{2}/\sqrt{m_{1}m_{2}}$.
\end{widetext}

Eq.~(\ref{Eq:DynMat}) can also be interpreted as an eigenvalue problem where the Bloch Hamiltonian matrix to diagonalize is given by Eq. (1) of the main text, repeated here 
\begin{equation}
{\bf H}_{q}=h_{0}\I+h_{z}\sigma_{z}+{\bf \bar{H}}_{q},\label{eq:qH_full}
\end{equation}
where $\sigma_z$ is Pauli $z$-matrix and $\I$ is the $2\times 2$ identity matrix; $h_{0}=(1/2)[(\omega_1^2+\omega_2^2)+(m_1+m_2)(K_1+K_2)/m^2]$ and $h_{z}=(1/2)[(\omega_1^2-\omega_2^2)+(m_2-m_1)(K_1+K_2)/m^2]$ are $q$-independent combinations of the lattice parameters. The matrix ${\bf  \bar{H}}_{q}$ is defined by
\begin{equation}
{\bf  \bar{H}}_{q}=-\frac{1}{m}\begin{bmatrix}0 & f^*(q)\\
f(q) & 0
\end{bmatrix},
\end{equation}
where $m=\sqrt{m_1 m_2}$ and $f(q)=K_1+K_2 e^{iq}$. One needs $h_0\ge 0$ and $h_0-|h_{z}|\ge 0$ for observing propagating modes ($h_0-|h_z|$ should be positive to guarantee propagating modes in the lower band). 

\section{Winding Number and Edge Modes} \label{A0}

The non-trivial topological nature of the lattice is most clearly seen by first setting $m_1=m_2=m$ and $\omega_1=\omega_2=\omega$, so that only the coupling constants alternate. Under these conditions, $h_z\equiv 0$ in Eq.~(\ref{eq:qH_full}) resulting in the matrix eigenproblem
\begin{equation}
{\bf \bar{H}}_{q}\ket{u_q}=\left[\omega_{q}^{2}-\omega^{2}-\frac{K_{1}+K_{2}}{m}\right]\ket{u_q},\label{eq:qHeig}
\end{equation}
where $|u_q\rangle=[u_{q,1},u_{q,2}]^{T}$.
In this form, the Hamiltonian resembles the Su-Schrieffer-Heeger (SSH) model of electrons hopping in polyacetylene~\cite{SSH79,AsbothBook,ShenBook}.
The function $f(q)=K_1+K_2 e^{iq}$ is periodic with period $2\pi$. 
Using the Pauli matrices $\sigma_{x,y}$, one can write ${\bf  \bar{H}}_{q}=R_x\sigma_x +R_y\sigma_y$, where $R_x=K_1+K_2\cos(q)$ and $R_y=-K_2\sin(q)$.
Interpreting the values of $(R_x,R_y)$ as a 2D vector, one may ask how many times does this vector wind around the origin when $q$ changes from $0$ to $2\pi$. It is easy to see that this quantity, usually referred to as the {\it winding number}, $\W$, evaluates to 
\begin{equation}
\W=\begin{cases}
1, & K_{1}<K_{2}\\
0, & K_{1}>K_{2}
\end{cases}\label{eq:wn}
\end{equation}
The winding number is an important topological property of a 1D system's band structure \cite{Hasan10,AsbothBook,Chiu2016} (or its dispersion relation for classical systems), encoded by its Hamiltonian. In what follows we will see that this number can be used to count the number of edge modes in a finite-length dimerized harmonic lattice. The Zak phase~\cite{Zak89} is the 1D Berry phase with the definition $\theta_Z=\int_0^{2\pi} dq\langle u_q |\partial_q\ket{u_q}$. For the dimerized lattice, $\theta_Z=2\pi\W$.

Up to now, the quasimomentum $q$ has assumed real values, yielding propagating Bloch modes. Another important class of modes can be obtained by (i) analytically continuing Eq.~\eqref{eq:qHeig} to complex $q$ and (ii) requiring that the amplitude of resulting normal modes vanish exactly on the second sublattice, i.e., $|u_{q}\rangle=[1,0]^{T}$. Eq.~\eqref{eq:qHeig} then reduces to two equations: $\omega_{q}^{2}=\omega^{2}+(K_{1}+K_{2})/m$ and $f(q)=0$. The first equation defines the frequency of the normal mode, which always lies in the gap between the two bulk bands. The second equation, which can be rewritten as $q=\pi-i\log (K_{1}/K_{2})$, implies that when $K_{1}\neq K_{2}$, the imaginary part of $q$ is
nonzero and the vibrational amplitude decays (grows) exponentially
as $n\rightarrow+\infty$ for $K_{1}<K_{2}$ ($K_{1}>K_{2}$).

We now truncate the dimerized lattice from the left by setting $m_n\rightarrow\infty$, $n=0,-1,-2,...$, or, equivalently, by the boundary condition $u_{q,0}=0$. The exponentially decaying mode ($K_1<K_2$) satisfies this boundary condition by construction and is normalizable in $n\in[1,\infty)$. This mode then becomes an {\it edge mode} of the truncated lattice. No edge mode exists for $K_1>K_2$ since the corresponding exponentially growing mode is non-normalizable. 
The winding number, Eq.~\eqref{eq:wn}, succinctly captures this condition. Thus, even though it is a rather abstract topological property, it nevertheless has a very important physical interpretation as it gives the number of edge modes in truncated dimerized lattices. 

Similar considerations apply to the case where the lattice is truncated from the right by setting $u_{q,N+1}=0$. Truncating the lattice from both sides produces a finite lattice of length $N$. Assuming that $N$ is large, left and right edge modes do not interact with each other and the number of left ($N_L$) and right ($N_R$) edge modes can be succinctly expressed as
\begin{align}
N_L&=\W,\nonumber\\
N_R&=\frac{1-e^{i\pi(N+\W)}}{2}.
\end{align}
As the total number of normal modes of the lattice of length $N$ has to be $N$, there is also $N-(N_L+N_R)$ delocalized modes. These modes of the finite lattice are obtained as linear combinations of propagating Bloch modes of the infinite lattice subject to the boundary conditions. Figure 1(c) of the main text shows the polarization vector squared of two edge (dashed red and solid blue lines) and one delocalized (dot-dashed green line) mode as a function of position along the lattice. The polarization vectors of the edge modes decay with exponent $\xi=-\log (K_{1}/K_{2})$. An index-theorem approach gives a way to count the edge modes for more generic cases~\cite{Kane14,Lubensky15}.

Due to their topological nature, the edge modes discussed above (alternating $K_n$, uniform $m_n$ and $\omega_n$), are expected to be robust to certain classes of perturbations/modifications. Figure 2(a-b) of the main text show the evolution of the edge modes (through the corresponding eigenvalues) as $K_{1}/K_{2}$ increases. The edge modes remain so long as their frequency is separated from the bulk phonon bands, which occurs in both the alternating coupling only case (Fig. 2(a)) and when other parameters alternate (Fig. 2(b)). This is also true for the presence of nonlinearity (to be discussed in a moment). 

Moreover, the edge modes remain even when masses and/or onsite frequencies alternate. Extending the SSH model to a time-dependent model backs the robustness of the edge modes in the presence of these additional alternating parameters~\cite{ShenBook}, as we will observe now in its classical analogue. By construction, each edge mode  vanishes exactly on one of the sublattices. Due to this, one can verify, by direct evaluations, that adding an alternating onsite potential corresponding to the $\sigma_z$ term in the Bloch Hamiltonian will not destroy the edge mode. However, the chiral (sublattice) symmetry is broken by the $\sigma_z$ term and as a consequence, the surviving edge modes are no longer pinned at zero frequency. Edge modes of the lattice with only alternating coupling constants are then also normal modes of the Hamiltonian with additional uniform or alternating onsite potentials, albeit with shifted frequencies.

When the total number of sites is odd and the edge mode starts from the first site, the frequency is $\omega_{b}=\sqrt{h_{0}+h_{z}+b_{1}}$. The gap between the edge mode and the
bottom of the upper band (or the gap between the edge mode and the top of the lower band) is \[ \sqrt{h_{0}+\frac{1}{2}\left(b_{1}+b_{2}+\sqrt{(h_z+b_{1}-b_{2})^{2}+4(d_{1}-d_{2})^{2}}\right)}-\omega_{b} \] (or  \[\omega_{b}-\sqrt{h_{0}+\frac{1}{2}\left(b_{1}+b_{2}-\sqrt{(h_z+b_{1}-b_{2})^{2}+4(d_{1}-d_{2})^{2}}\right)}\] ). Therefore, the edge mode always sits in between the two bands if $K_1\neq K_2$, $h_0 \ge 0$ and $h_0-|h_z|\ge 0$.
In the regime ($K_{1}<K_{2}$) with $h_0=h_z=0$ and when $m_{n}=m$ and $\omega_{n}=0$, the frequency gap, $\Delta\omega$,
between the edge modes and the top of the lower band (the bottom
of the upper band) is $\Delta\omega=\sqrt{(K_{1}+K_{2})/m}-\sqrt{2K_{1}/m}$
($\Delta\omega=\sqrt{2K_{2}/m}-\sqrt{(K_{1}+K_{2})/m}$) for $m_{n}=m$
and $\omega_{n}=0$. The frequency gaps, $\Delta\omega$, do not vanish until $K_{1}=K_{2}$,
after which the edge modes merge with the bulk. This is apparent from Fig.~2(a-b) of the main text.

\section{Rigorous derivation of Eq.~(6) of the main text}\label{A4}
We elucidate the origin of the topological component as an interfacial phenomenon, where the edge modes deplete the coupling of the bulk modes to the external sources/sinks of thermal energy. To rigorously derive Eq. (6) in the main text, we use Dhar's modified version of Eq.~\eqref{eq:CLfull} \cite{Dhar2001} (with symmetric couplings to the reservoirs, $\gamma=\gamma_L=\gamma_R$),
\begin{equation}
\kappa=\frac{k_B}{\pi}\gamma^2 \int^{\infty}_{-\infty} d\omega \omega^2 |Y^{-1}_{1N}(\omega)|^2,
\label{eq:Dhar}\end{equation}
where $Y=\Phi-\omega^2 M-A$, with $\Phi$ defined above, $M_{i,j}=m_i \delta_{i,j}$ is the mass matrix, and $A_{i,j}=-i \gamma \omega \delta_{i,j}(\delta_{i,1}+\delta_{i,N})$ represents the interaction with the Langevin reservoirs.
Writing the normal modes as $|q\rangle=\sum_i u_{qi}|i\rangle$, the $Y$-matrix takes on a simple form,

\begin{equation}
Y=\sum_q  (w^2-w_q^2)|q\rangle\langle q|-i \gamma \omega \sum_{q,q^\p } \left( u_{q1} u_{q^\p 1} +u_{qN} u_{q^\p N} \right) |q\rangle \langle q^\p | .
\end{equation}
For small $\gamma$, the terms with $q^\p =q$ are dominant in $Y^{-1}$, which then satisfies $Y^{-1}\approx \left[  w^2-w_q^2-i \gamma \omega  \left( u_{q1}^2 +u_{qN}^2 \right)  \right]^{-1} |q\rangle\langle q|$. Substituting this into Eq.~\eqref{eq:Dhar}, gives
\begin{equation}
\kappa\approx\frac{k_B}{\pi}\gamma^2 \int^{\infty}_{-\infty} d\omega \omega^2 \frac{u_{q1}^2 u_{qN}^2  }{ \left| w^2-w_q^2-i \gamma \omega  \left( u_{q1}^2 +u_{qN}^2 \right) \right|^2} .
\label{eq:Dhar2}\end{equation}
Evaluating by integrating over the poles and taking the limit of small $\gamma$ leads to Eq. (6).

We apply this equation heuristically. To do so, we need the amplitudes of the modes at each boundary. These are found in the standard way, first by ``mass-weighting" the coordinates (i.e., multiplying by $\Op{M}^{1/2}$, with $\Op{M}$ the diagonal matrix of masses) and then via an orthogonal transformation $\Op{T}$ that diagonalizes the weighted matrix of coupling constants (including onsite frequencies), $\Op{M}^{-1/2} \Op{K} \Op{M}^{-1/2}$. The amplitudes are $u_{qn}=\left[\Op{M}^{-1/2}\Op{\Op{T}} \right]_{nq}$. Instead of solving exactly for these amplitudes (which is tantamount to solving the full problem, as done above), we estimate them by exploiting properties of orthogonal transformations, namely, that $\sum_q u_{qn}^2 \equiv 1/m_n$. When there is a highly localized mode, i.e., one with a large amplitude at a particular site, it means that all other modes must have a substantially reduced amplitude on that site. This is the case when edge modes are present, giving the simple equation $\sum_{q\in\Omega}u_{qn}^{2}=1/m-\sum_{q\in\SE}u_{qn}^{2}$.
To get an estimate of $u_{q1}^2$ for $q\in\Omega$, we just need $u_{q1}^{2}$ for $q\in\SE$. Along the lattice, a left edge mode has amplitude proportional to $(1,0,(-K_{1}/K_{2}),0,(-K_{1}/K_{2})^{2},\cdots)$. Taking the infinite lattice limit $N\rightarrow\infty$, the normalization factor without the mass scaling is $\sqrt{1-(K_{1}/K_{2})^{2}}$. This gives $u_{q1}^2=\left(1-(K_{1}/K_{2})^{2}\right)/m_1$ for the edge mode. Thus, an estimate of the amplitude squared for $q\in\Omega$ is $u_{q1}^2 \approx (K_{1}/K_{2})^{2}/ N m_1$, where we divide by $N$ because there are $\approx N$ bulk-like modes. This is the expression we use in the main text.

\section{Rigorous derivation of Eqs.~(10) and (11) of the main text}\label{A3}

When a thermal current is driven through the lattice by contact with thermal reservoirs at the boundaries, the equation of motion, Eq.~\eqref{eq:EOM1}, becomes
\begin{eqnarray}\label{eq:EOM2}
m_n\ddot{x}_n&=&-(m_n \omega_n^2+K_{n-1}+K_{n})x_n +K_{n-1}x_{n-1}+K_{n}x_{n+1} + \gamma_n\dot{x}_n+\eta_n,
\end{eqnarray}
where $\gamma_n = (\gamma_L \delta_{n,1}+\gamma_R \delta_{n,N})$ and $\eta_n = \eta_L \delta_{n,1}+ \eta_R \delta_{n,N}$ for Langevin reservoirs \cite{Dhar2008,Lepri_review}. The left ($L$) and right ($R$) reservoirs have temperatures $T_L$ and $T_R$, respectively, where the temperature difference, $\Delta T=T_{L}-T_{R}$, gives rise to a thermal current. The random forces, $\eta_{L,R}$, satisfy the fluctuation-dissipation theorem, $\langle\eta_{L,R}(\omega)\eta_{L,R}(\omega^{\prime})\rangle=4\pi\gamma_{L,R} T_{L,R}\delta(\omega+\omega^{\prime})$.

There are different but equivalent ways to obtain the conductance $\c$. Often, one computes the incoming thermal current at one of the boundaries, i.e., $J=\langle (\gamma_L\dot{x}_1+\eta_L)\dot{x}_1 \rangle$. Then, following Refs.~\cite{CL71,Dhar2001}, one performs a Fourier transform and solves the equations of motion. In the steady state,  the conductance is
\begin{eqnarray}\label{eq:CLfull}
\kappa &=& \frac{J}{\Delta T} = k_B\gamma_L\gamma_R\int_{-\infty}^{\infty}\frac{d\omega}{\pi}\omega^2|C_{1N}(\omega)|^2 \times   [(K_{1,N}-\omega^2\gamma_L\gamma_RK_{2,N-1})^2+\nonumber \\&+& \omega^2(\gamma_LK_{1,N-1}+ \gamma_RK_{2,N})^2]^{-1},
\end{eqnarray}
where  $ k_B $ is Boltzmann's constant and $C_{1N}$ denotes the cofactor of ${\Phi}_{1N}$. The matrix ${\Phi}$ has elements ${\Phi}_{n,n^{\prime}}=(m_n \omega_n^2+K_{n-1}+K_{n})\delta_{n,n^{\prime}}-K_{n-1}\delta_{n,n^{\prime}-1}-K_{n}\delta_{n-1,n^{\prime}}$ for $n,n^{\prime}=1,\cdots,N$. $K_{i,j}$ denotes the determinant of ${\Phi}$ starting from the $i$-th site and ending with the $j$-th site.

For a uniform lattice in the infinite-length limit, analytic expressions for $\kappa$ were already found in Ref.~\onlinecite{CL71}. For a periodic lattice with a two site unit cell, analytic expressions can also be found for arbitrary $\gamma_L$ and $\gamma_R$ by generalizing the formalism of Ref.~\onlinecite{ChienExactDiag16}. However, for $\gamma_L = \gamma_R = \gamma$ and in the small $\gamma$ regime, the conductance in the infinite-length limit simplifies to
\[
\c  = k_B\frac{2 \gamma}{\pi} \frac{K_1}{K_2} \int_{\Omega}\frac{d\omega |\omega\sin(q)|}{|\alpha_1+\alpha_2|} \label{eq:EvenExact}
\]
for an even length lattice and
\[
\c = k_B\frac{2\gamma}{\pi}\int_{\Omega} \frac{ d\omega |\omega\sin(q)|}{\left| \alpha_1\left(\frac{K_1}{K_2}+\frac{K_2}{K_1} \right) \right|}  \label{eq:OddExact}
\]
for an odd length lattice.
The integrals are over the bands $\Omega$, $\alpha_1=m_2 \omega_2^2+K_1+K_2-m_1\omega^2$, $\alpha_2=m_2 \omega_2^2+K_1+K_2-m_2\omega^2$, and $q$ satisfies $\alpha_1\alpha_2=K_1^2+K_2^2+2K_1K_2\cos(q)$. While these expressions are from a small $\gamma$ expansion, they often work up to moderate values of $\gamma$ (order 1 or larger in $\sqrt{m_2 K_2}$), as the presence of edge modes effectively weakens the coupling (and the zero edge mode case has no anomalous behavior).

Both integrals are analytically tractable. However, Eq.~\eqref{eq:EvenExact} immediately yields Eq. (10) in the main text for the even site lattice, as going from the nontrivial topology with two edge modes to the trivial one with no edge modes is a swap of $K_1$ and $K_2$. Since $\alpha_1$ and $\alpha_2$ both involve $K_1+K_2$, the only effect of the swap is to introduce a factor $(K_2/K_1)^2$. Using the relation between the correlation length and the coupling constants, $\xi=-\log (K_{1}/K_{2})$, gives Eq. (10) in the main text. Moreover, one can swap masses and onsite frequencies in unison with the nearest-neighbor couplings, giving rise to the same result.

To perform the integrals, one changes variables to momentum, obtaining a form
\[
\frac{1}{\pi} \int dq \frac{A \sin^2 q}{B+C \cos q}
\]
for both cases. Integrating separately over each band and adding the results yields
\[
\frac{2 A}{B+\sqrt{B^2-C^2}}.
\]
The quantities in the expression are $A=2 k_B \gamma K_1^2 (m_1+m_2)$, $B=(D_2^\p m_1 - D_1^\p m_2)^2+(m_1+m_2)^2(K_1^2+K_2^2)$, and $C=2 K_1 K_2 (m_1+m_2)^2$ for the even length lattice, where $D_i^\p=m_i \omega_i^2 + K_1 + K_2$.
While complicated, the resulting equation factorizes the topological component, given by $\varXi$ of Eq. (11), from the non-topological component, as only the factor $K_1^2$ in $A$ has those coupling constants appear asymmetrically. Moreover, one need not directly use this result, but rather simply take the ratio of the conductance of the lattice to that with the parameters swapped.

For the odd length lattice, we have $A=2 k_B \gamma K_2^2 / m_1 (1+K_2^2/K_1^2)$, $B=K_1^2+K_2^2$, and $C=2 K_1 K_2$. This yields the conductance
\[
\c = \frac{k_B \gamma}{m_1 \left( 1 + K_2^2/K_1^2 \right)},
\]
which is an exact expression for the small $\gamma$ regime. This gives Eq. (10) of the main text for the odd length lattice. 

\section{Nonlinear FPU lattices}\label{A5}
We develop the self-consistent phonon theory for generalized Fermi-Pasta-Ulam model \cite{Burkalov1990,James2007,Livi1997,Aoki1992}, in which the nonlinear coefficients also alternate between two values for odd and even $n$. The full Hamiltonian is given by
\[
H = \sum_{n=1}^{N}\frac{m_{n}}{2}\left(\dot{x}_{n}^{2}+\omega_{n}^{2}x_{n}^{2}\right)+ \sum_{n=0}^{N} V(x_{n}-x_{n+1}),
\]
where the nearest-neighbor potential is
\begin{eqnarray}
V(x_{n}-x_{n+1}) &=& \frac{K_{n}}{2}(x_{n}-x_{n+1})^{2}+\frac{\alpha_{n}}{3}(x_{n}-x_{n+1})^{3}  + \frac{\beta_{n}}{4}(x_{n}-x_{n+1})^{4}.  \label{eq:Hfpu}
\end{eqnarray}
The parameters $m_n, \omega_n, K_n, \alpha_n$ and $\beta_n$ all alternate as a function of $n$.  The coordinates at the boundaries are fixed, i.e., $x_0 = x_{N+1} = 0$.

To proceed, we follow the self-consistent phonon approximation \cite{Cao2015,He2008,Werthamer1970}, in which the FPU Hamiltonian is mapped onto a harmonic lattice Hamiltonian
\begin{equation}
H_0=\sum_{n=1}^{N}\frac{m_{n}}{2}\left(\dot{y}_{n}^{2}+\omega_{n}^{2}y_{n}^{2}\right)+ \sum_{n=0}^{N} \frac{f_{n}}{2}(y_{n}-y_{n+1})^{2},
\end{equation}
for which we define $\Delta x_n = x_n-x_{n-1}\equiv y_n-y_{n-1}+l_n$, where $l_n=\langle (x_n-x_{n-1}) \rangle_0$ is the average inter-particle distance. The self-consistent phonon approximation requires that
\begin{eqnarray}
\left\langle \frac{\partial V}{\partial \Delta x_n} \right\rangle_0 & =& 0 \nonumber \\
\left\langle \frac{\partial^2 V}{\partial \Delta x_n^2} \right\rangle_0 & =& f_n,
\end{eqnarray}
where
\begin{equation}
\langle A({\bf y}) \rangle_0=\frac{\int A({\bf y}) \exp\left(-H_0({\bf y}) /k_B T \right) d{\bf y}} {\int \exp\left(-H_0({\bf y}) /k_B T \right) d{\bf y}} .
\end{equation}
Directly evaluating the Gaussian integrals leads to the self-consistent equations
\begin{eqnarray}
K_n l_n +\alpha_n (l_n^2+\langle \Delta x_n^2\rangle_0)+\beta_n (l_n^3+3l_n \langle \Delta x_n^2\rangle_0 ) &=& 0 \nonumber \\
K_n  +2 \alpha_n l_n+3 \beta_n (l_n^2+\langle \Delta x_n^2\rangle_0 ) &=& f_n,
\end{eqnarray}
with (for uniform masses)
\begin{equation}
\langle \Delta x_n^2\rangle_0 = \frac{k_B T}{m} \sum_q \frac{\left(\Op{T}^{-1}_{q,n}-\Op{T}^{-1}_{q,n-1}\right)^2}{\omega^2_q},
\end{equation}
where $\Op{T}$ is the orthogonal transformation that diagonalizes the lattice (mass-weighting is not necessary for uniform masses) and $\omega_q$ are the eigen-frequencies. Since the system is finite, $\Op{T}_{q,0}=\Op{T}_{q,{N+1}}=0$ for all $q$. These equations are solved numerically to find the self-consistent phonon modes. 

\section{Details of Figure 3}\label{A6}
Figure 3(a) of the main text is the conductance versus $K_{1}/K_{2}$ in the $N \to \infty$ limit for $\gamma=10^{-4} \sqrt{m_2 K_2}$. We use expressions for the conductance that are exact for all $\gamma$. Solid lines are for $K_1$ and $K_2$ alternating but otherwise a uniform lattice. Dashed lines are for $m_1 = 2 m_2$ and $\omega_1 = \omega_2$. Dashed-dotted lines are for $m_1/m_2 = K_1/K_2$ and $\omega_1 = \omega_2$. For each of these parameter sets, four lines are shown in different colors (purple shows the two edge mode lattice, green the zero edge mode lattice, and red and blue show the lattice with one edge mode on the left and right, respectively; to swap the side of the edge mode, all parameters are swapped). Figure 3(b) takes the ratios as described in the main text. Figure 3(c) shows three calculations, one with nonlinearity and two with disorder.

When disorder is present, a 1D system always undergoes Anderson localization \cite{Anderson1958}. Consequently, if the lattice length is larger than the localization length (which depends on the strength of disorder), the eigenmodes are exponentially localized, which will lead to an exponential reduction of the thermal conductance, obscuring the effect of the edge modes. In the calculation we take mass and nearest neighbor couplings to be random variables taken from uniform distributions. The disorder strength is 5 \% of the parameters' value, which is such that for $N=32$ ($N=33$ for the odd length lattice) the bulk states are not yet localized, and only when $K_1$ nears $K_2$ (i.e., close to $K_1/K_2=1$) does disorder play a role -- in this regime, the edge modes disappear and disorder becomes the primary factor in reducing the conductance. The conductance is averaged over 1000 realizations of disorder and the associated error bars are smaller than the symbol.  The coupling to the Langevin reservoirs is $\gamma=10^{-2} \sqrt{m_2 K_2}$. Yellow circles are for disorder in only $K_1$ and $K_2$ on top of $m_1 = 2 m_2$ and $\omega_1 = \omega_2$. Wine squares are for disorder in both masses and nearest-neighbor coupling constants on top of $m_1 = 2 m_2$ and $\omega_1 = \omega_2$. As with the infinite length lattice, we compute the conductance exactly (i.e., not assuming small $\gamma$), which here means we use the procedure of Ref.~\cite{velizhanin15}.

The nonlinearity calculation (magenta diamonds) is for the $\beta-FPU$ lattice in Fig.~2(c) and uses Eq.~\eqref{eq:Dhar} with the self-consistent modes. We change the ratio $K_1/K_2$, but we keep all other parameters (including $\beta_1, \beta_2$) fixed. The coupling to the Langevin reservoirs is $\gamma=10^{-2} \sqrt{m_2 K_2}$. Since the self-consistent calculation of the couplings in a finite lattice introduces both correlated disorder and a systematic deviation of parameters, an effective ratio $K_1/K_2$ needs to be determined, which is done with the following procedure. For the odd lattice with an edge mode on the left, the effective $K_1/K_2$ is determined via $\left( K_1 / K_2 \right)^2=1-u_{q1}^2+u_{qN}^2$,
where $q \in \E$ is the edge mode (see discussion below Eq.~\eqref{eq:Dhar2}) and this expression includes a finite system correction ($+u_{qN}^2$, which properly normalizes the exponentially decaying mode on the finite lattice for the odd length case). For the even case, the effective $K_1$ and $K_2$ are taken as the average couplings in the even and odd positions, respectively.

\section{Potential experimental realization}
The effect we predict is robust and detail-independent, and large arrays (e.g., 100 sites) of MEMS and NEMS devices with well defined vibrational dynamics~\cite{Buks2002,Sato2003,Krylov2014} will be ideal for measuring the topological suppression of energy transport. These lattices can be fabricated using conventional lithographic and reactive ion etching techniques, and the device layer properties and geometries can be artificially tailored to achieve natural harmonics ranging from hertz to gigahertz~\cite{Spiel2001,Huang2003,Mitcheson2004,Liu2011}. To elucidate the dynamics of energy transport, a fabricated lattice of cantilever beams can be placed into a vacuum chamber to eliminate viscous damping effects (although this may not be needed for high enough frequency resonances).  Thermal excitation can be achieved using a laser source or Joule heating via a fabricated serpentine metal electrode~\cite{Zalalutdinov2003,Ilic2005} and motion detection of the cantilevers can be accomplished either optically or electro-statically~\cite{Nathanson1965,Meyer1988,Carr1997}. The dynamics of the impinging thermal energy can be further tailored by introducing local constraints within the lattice (e.g., beam architectures with individually addressable electrodes). This scenario will present a direct visualization of energy transport using dynamics of coupled micro- and nano-mechanical resonators.

\bibliography{reference1}

\begin{figure}[b]
\includegraphics{Inv.png}
\end{figure}